\documentclass[aps,pra,tightenlines,12pt,superscriptaddress,notitlepage,nofootinbib,longbibliography]{revtex4-1}

\usepackage{graphicx}  
\usepackage{xcolor}
\usepackage{dcolumn}   
\usepackage{bm}        
\usepackage{amssymb}   
\usepackage{amsmath}
\usepackage{verbatim}
\usepackage[hidelinks]{hyperref}
\usepackage[utf8]{inputenc}     
\usepackage{tabularx}
\usepackage{multirow}
\usepackage[capitalise,noabbrev]{cleveref}
\usepackage[centering,margin=1in]{geometry}
\usepackage{braket}
\usepackage{setspace}

\AtBeginDocument{%
    \newwrite\bibnotes
    \def\bibnotesext{Notes.bib}
    \immediate\openout\bibnotes=\jobname\bibnotesext
    \immediate\write\bibnotes{@CONTROL{REVTEX41Control}}
    \immediate\write\bibnotes{@CONTROL{%
    apsrev41Control,author="08",editor="1",pages="1",title="0",year="1"}}
     \if@filesw
     \immediate\write\@auxout{\string\citation{apsrev41Control}}%
    \fi
}%

\newcommand\snowmass{\begin{center}\rule[-0.2in]{\hsize}{0.01in}\\\rule{\hsize}{0.01in}\\
\vskip 0.1in Submitted to the  Proceedings of the US Community Study\\ 
on the Future of Particle Physics (Snowmass 2021)\\ 
\rule{\hsize}{0.01in}\\\rule[+0.2in]{\hsize}{0.01in} \end{center}}

\begin{document}

\title{Snowmass 2021 White Paper: \\ Observational Signatures of Quantum Gravity}

\author{Kathryn M. Zurek}
\affiliation{Walter Burke Institute for Theoretical Physics\\California Institute of Technology, Pasadena, CA }

\setstretch{1.2}

\begin{abstract}

This short review is intended as a colloquium-level summary, for the Snowmass 2021 process, on recent theoretical results on infrared observables in quantum gravity.  
We rely on simple physical arguments, most notably a random walk intuition, to show how effects of quantum gravity in the ultraviolet (at the Planck length $\ell_p \approx 10^{-35} \mbox{ m}$) may integrate into the infrared when the large measurement length scale $L$ enters into the observable.  A quantum uncertainty at lightsheet horizons would give rise to an accumulated effect of size $\delta L^2 \simeq \ell_p L/4 \pi$.  We discuss how the random walk intuition falls out from more formal calculations, such as from AdS/CFT, from the dimensional reduction of the Einstein-Hilbert action to dilaton gravity, from multiple gravitational shockwaves generated by vacuum energy fluctuations, as well as from an effective description of gravity as a fluid.  We overview experimental prospects for measuring this effect with a simple Michelson interferometer utilizing many of the tools developed for gravitational wave observatories.  

\end{abstract}

\snowmass

\maketitle

\newpage

\tableofcontents

\section{Introduction}

From an Effective Field Theory point of view, effects of Quantum Gravity should appear at length and time scales set by the Newton constant:
\begin{equation}
\ell_p = \sqrt{8 \pi G \hbar/c^3} \simeq 10^{-34}\mbox{ m},~~t_p = \ell_p/c \simeq 10^{-45} \mbox{ s},
\end{equation}
where $\ell_p,~t_p$ are the Planck length and time, respectively.  Clearly these length and time scales are far too small to be observed.  This simple argument is one of the main reasons for the conventional view that the quantum effects of gravity are not observable.  One simple way to think about this intuition is by computing loop effects of gravitons on the Newton potential:
\begin{equation}
V(r)  = - \frac{G m_1 m_2}{r} \left( 1 + a \frac{G(m_1 + m_2)}{r} + b \frac{G \hbar}{r^2 c^3} \right).
\end{equation}
See Ref.~\cite{Donoghue} for lecture notes introducing quantum gravity as an EFT.

\begin{figure}[t]
\includegraphics[scale=0.4]{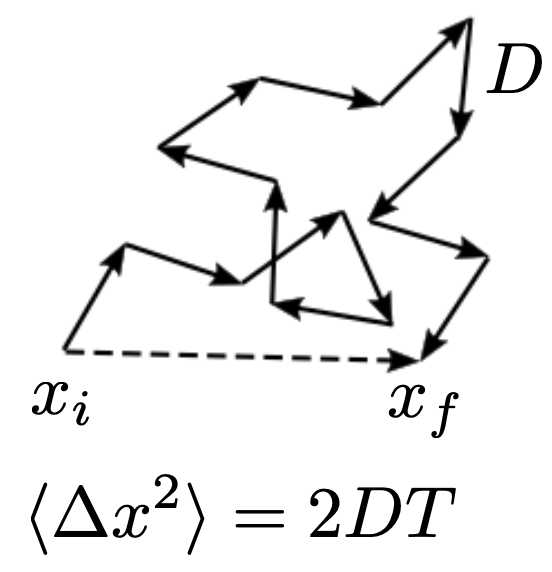}
\hspace{0.5in}\includegraphics[scale=0.4]{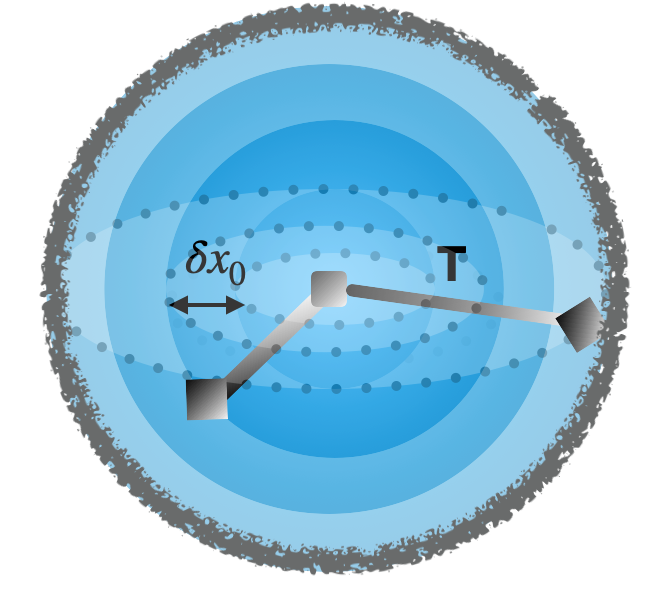}
\caption{{\em left}: Schematic of the random-walk intuition.  If a particle starts at some position $x_i$, the variance of its final position $x_f$, $\langle (x_i - x_f)^2 \rangle = 2 D T$ will be set by the diffusion coefficient $D$ and the amount of time that elapsed between initial and final measurement.  The diffusion coefficient is in turn determined by the typical time separation $D \sim \delta x_0$ between interactions that randomize the direction of the particle's trajectory.  This random-walk intuition works well to describe formal results in the blurring of one of the lightcone directions in the causal development of light sheets.  {\em right}: The causal development of spherical light sheets over a time period $T$, such as occurs in an interferometer having two mirrors at an equal distance $L = T$ from the beamsplitter at the center of the sphere.  Over the course of this whitepaper, we show that there is an analogy between the random walk and the causal development of a region that undergoes (statistically uncorrelated) shocks separated by a distance $\delta x_0 = D$.  The integrated uncertainty can be thought of as a blurring of the horizon (shown as the fuzzy line outlining the spherical shells) of size $\delta L^2 = 2 D L$. We will be able to derive this result formally from multiple theoretically distinct, but physically equivalent, calculations.  Right panel taken from Ref.~\cite{BZ}.}
\label{figure-RW}
\end{figure}

However, we know that in some cases, observable effects can appear at long distances in comparison to the microscopic interaction scales.   A classic example is diffusion; see the illustration in Fig.~\ref{figure-RW}.  In diffusion, even though the interactions are purely local, the uncertainty in a particle's position grows, with the total effect spreading itself non-locally over time.    Specifically, consider a gas where the typical inter-particle spacing is $R$, and the velocity of the particles is $v$ such that the typical time between collisions is $\delta x_0 = R/v$.  If we consider the position $x$ of some typical particle, as time progresses, the uncertainty on its position, $\Delta x = x_f - x_i$, will increase due to diffusion:
\begin{equation}
\langle \Delta x^2 \rangle = 2 D T,
\end{equation}
where $T$ is the amount of time that has elapsed since the particle was initially localized at some position $x_i$.  Importantly, the diffusion constant $D$ is set by a microsopic scale, $D \sim v^2 \delta x_0$, which is directly related to the particle mobility, or the typical length or time scale between particle interactions.  Note that $D$ need not depend on an integer power of the coupling constant governing the interaction, such that the uncertainty in $\Delta x$ also need not depend on an integer power of the coupling constant, violating our EFT intuition.  This is possible, in part, because the effect accumulates over time, {\em i.e.} there is a memory effect from the random walk behavior:
\begin{equation}
\langle \Delta x^2 \rangle \sim N \delta x_0^2,
\end{equation}
where $N = T/\delta x_0$ is the number of steps in the random walk.  This has the feature of ``root-N'' statistics
\begin{equation}
\Delta x \sim \sqrt{N} \delta x_0 \sim \sqrt{T \delta x_0},
\label{eq:RW}
\end{equation}
with fluctuations in the particle's position $\Delta x$ dependent on {\em both} the ultraviolet (UV) interaction scale $\Delta t$ and the number of interactions $N$.  Note that the effect of the UV scale is effectively transmuted into a much larger effect at longer distance scales via the large number of interactions.  This diffusion behavior, rather famously, is an important observable effect that led to the development of statistical mechanics.

We will refer to this effect as ``random walk intuition,'' but our goal is to see whether there is reason to believe, more formally, that such effects could occur within Quantum Gravity, utilizing well-studied toy systems, {\em i.e.} we seek to determine if the scaling in Eq.~\ref{eq:RW} appears naturally in more formal contexts, utilizing the properties of causal diamonds.   But let us first introduce the system of interest and make the analogy with the the random walk intuition.  We are interested in spacetime  near horizons created by the causal development of light sheets; to measure the geometry of spacetime, we will use an interferometer.  The interferometer is shown in the right panel of Fig.~\ref{figure-RW}, where the causal development of the spherical region probed by the interferometer arms is represented by concentric spheres.  The analogue of $\delta x_0$ in the random walk is the typical time scale for the metric to experience an ${\cal O}(1)$ perturbation; in four dimensions, we expect this timescale to be Planckian.  The important point is whether the spacetime fluctuations can {\em accumulate} along the lightcone directions. 
The accumulated effect of the metric perturbations along the lightcone directions will present itself as a quantum uncertainty in the position of the horizon, which is measurable if the two arms of the interferometer experience slightly different perturbations along their trajectories, analogous to waves.    

While this random walk intuition will turn out to be surprisingly effective to describe certain aspects of the physical effect, we seek to answer whether the effect arises formally in known theories of quantum gravity.  We will base our discussion on a series of works \cite{VZ1,VZ2,pixellon,BZ,BKZ,GLZ,VZ3}.  The first of these references proposed this effect (which we will sometimes call the VZ effect) based on a dictionary between black hole horizons and causal diamonds in the vacuum that is known to hold in certain formal contexts, like AdS/CFT \cite{VZ2}; they showed that the quantum uncertainty at horizons does in fact obey this random walk intuition.  Subsequent works, such as Refs.~\cite{BZ,BKZ,GLZ}, have argued that the more formal contexts in which these results can be derived can be extended to Minkowski space in four dimensions.  Over the next sections our goal is to outline these developments and make the case that an effect could be observable.  We will attempt to balance our presentation between results which can formally be shown and raw physical intuition (which turns out to be formally fairly correct). We will also discuss an experiment moving forward in the initial stages, GQuEST (Gravity from the Quantum Entanglement of SpaceTime), which could measure such fluctuations.

\section{Quantum Gravity at Horizons}

We are ultimately interested in measurements of the geometry of empty spacetime.  At first glance, it would seem that the diffusion effect would be irrelevant, because what, after all, would replace the particles whose density fixes the interaction time?

On the other hand, empty spacetime, when quantum mechanics is taken into consideration, is expected to be full of quantum fluctuations of the vacuum state.  There is a history of this idea, introduced by Wheeler, in the form of spacetime ``foam.''  We won't be interested in foam, as we are interested in more conventional quantum fluctuations of the vacuum that perturb the geometry of spacetime.   We will work in the framework of effects appearing, for example, in AdS/CFT, where quantum effects can, in some cases, be calculable.  We will argue that these features are {\em physically generic}, conform with the random walk intuition presented in the introduction, and apply outside the formal context of AdS/CFT, which we will largely regard as a calculational tool.

We are interested in quantum fluctuations at a horizon.  As discussed briefly in the introduction, a horizon is simply a surface that bounds a region that is in causal contact from a region that is not in causal contact.  The types of horizons that have been studied most extensively in the context of  quantum gravity are black hole horizons, about which many very interesting things have been discovered.  The first is that, although the horizon of a black hole simply passes through an empty region of spacetime, it has an entropy associated with it:
\begin{equation}
S_{BH} = \frac{A}{4 G} = S_{\rm ent},
\label{eq:BH}
\end{equation}
where the subscripts denote Bekenstein-Hawking (BH) and entanglement (ent), and we have equated these two entropies.  The second equality is known to be true only in certain systems (such as in AdS/CFT), where the BH entropy exactly gives the entanglement entropy of the boundary quantum field theory, see for example \cite{RT}.  For reasons we will discuss below, there is evidence that the second equality holds more generally.  One intuitive way to understand the equality in the context of black holes is as follows.  Entropy has information associated with it, and the entropy counts the degrees of freedom at the horizon.  It has now been understood that entanglement in these degrees-of-freedom at the black hole horizon play a crucial role in learning how information is retrieved from a black hole as it evaporates, where a paradox between quantum mechanics and gravity is sharpened.  What happens to information as a black hole evaporates, if it cannot escape (by locality), and it cannot be destroyed (by unitarity)?   Non-locality and entanglement in quantum mechanics at long distances play a key role, not only the Planck length.   There is also evidence for the equality Eq.~(\ref{eq:BH}) more generally for any quantum field theory restricted to a causal diamond {\em e.g.} \cite{CHM,Jacobson:2015hqa,jv}. 

Taking the idea seriously that the Bekenstein-Hawking entropy, Eq.~(\ref{eq:BH}), gives the number of quantum degrees of freedom that can fluctuate, leads to the idea of the quantum width of a black hole horizon.  This was discussed in a paper by Marolf \cite{Marolf}, who asked how perturbations in those degrees of freedom with temperature $T \sim 1/L$ given by the size of the black hole $L$ could change the position of the black hole horizon.  If we re-cast his solution, we find the quantum uncertainty in the position of the black hole horizon
\begin{equation}
\delta L^2 \sim \frac{L^2}{\sqrt{S_{BH}}},
\label{eq:Marolf}
\end{equation}
which says that the variance in the black hole horizon size is a $1/\sqrt{S_{BH}}$ effect.  Because the entropy appears in the expression, both the UV (Planck length $l_p$) and IR (black hole size $L$) appear.  This is also consistent with the ``root-N'' quantum error on a system having $S$ bits.  Specializing to $d = 4$, we obtain 
\begin{equation}
\delta L \sim \sqrt{l_p L}.
\label{eq:Marolf4d}
\end{equation}
Given the ``root-N'' quantum errors of Eq.~(\ref{eq:Marolf}), it is not surprising that the result agrees with the random walk intuition given by Eq.~(\ref{eq:RW}).  Though perhaps it is remarkable that the quantum uncertainty in a black-hole horizon would agree with that for an empty volume of spacetime measured by an interferometer, as shown in Fig.~\ref{figure-RW}.

The quantum degrees of freedom at horizons enter front and center into questions of holography and quantum gravity.  First, that the information of a black hole is bounded by an area, and not a volume, indicates that EFT vastly overcounts the degrees of freedom of spacetime.  Second, the black hole system has taught us that naive EFT reasoning breaks down at such holographic horizons, otherwise the contradiction between information preservation and black hole evaporation would not exist.    Lastly, entanglement between the degrees of freedom -- inside and outside the horizon -- is important.  

These ideas have been shown to hold for horizons more generally than black holes.  For example, in quantum field theories with massless degrees of freedom generally, the density matrix of the ground state can be traced over the degrees of freedom residing outside an imaginary sphere, giving a UV divergent entanglement entropy proportional to the area, and having the form Eq.~(\ref{eq:BH}) \cite{Srednicki:1993im}.  More specifically, tracing out the complement of a region bounded by some entangling surface gives rise to a thermal density matrix characterized by a ``modular Hamiltonian,'' and the entanglement entropy is given by the area of the entangling surface \cite{CHM,RT}.  Such a thermal density matrix will give rise to quantum fluctuations at horizons.  
Finally, the entanglement entropy of causal diamonds in empty space can sometimes be computed with Euclidean methods \cite{Sussug,Callan:1994py,Cooperman:2013iqr,bdf}, giving the result Eq.~(\ref{eq:BH}). 

Taken together, one begins to think that Eq.~(\ref{eq:BH}) is rather generic to horizons, and that the density matrix of the degrees of freedom within a horizon is thermal with a temperature given by the inverse size of the horizon.  The evidence has accumulated to such an extent that one can draw up a dictionary between a black hole horizon and a causal diamond, as shown in Table~\ref{table:dictionary}.    (The causal development of a horizon is known as a causal diamond.)  The dictionary can be explicitly demonstrated in certain contexts, such as for empty Ryu-Takayanagi diamonds in AdS/CFT, as we will discuss in the next section.  We will argue that the dictionary applies more generally, taking the dictionary seriously for empty causal diamonds in flat space.  Our purpose is to extend the notion of holography to flat space in quantifiable ways, utilizing the entropy associated with light sheet horizons and describing the quantum fluctuations in the geometry of flat spacetime in a stochastic language.  This will allow us to compute coarse-grained observables of quantum fluctuations of the geometry in a causal diamond.  Such quantum fluctuations are large enough to be within reach of experiment.  

\begin{table}
\caption{Black Hole - Causal Diamond Dictionary}
\centering
\begin{tabular}{|l || l  l|}
\hline \hline
\multicolumn{1}{c}{Black Hole} & \multicolumn{2}{c}{Casual Diamond} \\ \hline \hline
Horizon & Null sheets & \\ \hline
Horizon Area $A$ & Entangling surface area & $A$  \\ \hline
Black hole temperature $T$ & Size of causal diamond & $T \sim 1/L$ \\ \hline
Black hole mass $M$ & Modular fluctuation & $M = \frac{1}{2 \pi L} \left( K - \langle K \rangle \right)$ \\ \hline
Thermodynamic free energy $F_\beta$ & Partition function & $F_\beta = - \frac{1}{\beta} \log \mbox{ tr} \left(e^{-\beta K} \right)$ \\ \hline
Thermodynamic Entropy & Entanglement Entropy & $S = \frac{A}{4 G_N}$ \\ \hline
Thermodynamic entropy fluctuations & Modular Fluctuations & $\langle \Delta K^2 \rangle  = \frac{A}{4 G_N}$ \\ 
\hline
\end{tabular}
\label{table:dictionary}
\end{table}

\section{Measurements of Quantum Gravity at a Light-Sheet Horizon}

\begin{figure}[t]
\includegraphics[scale=0.4]{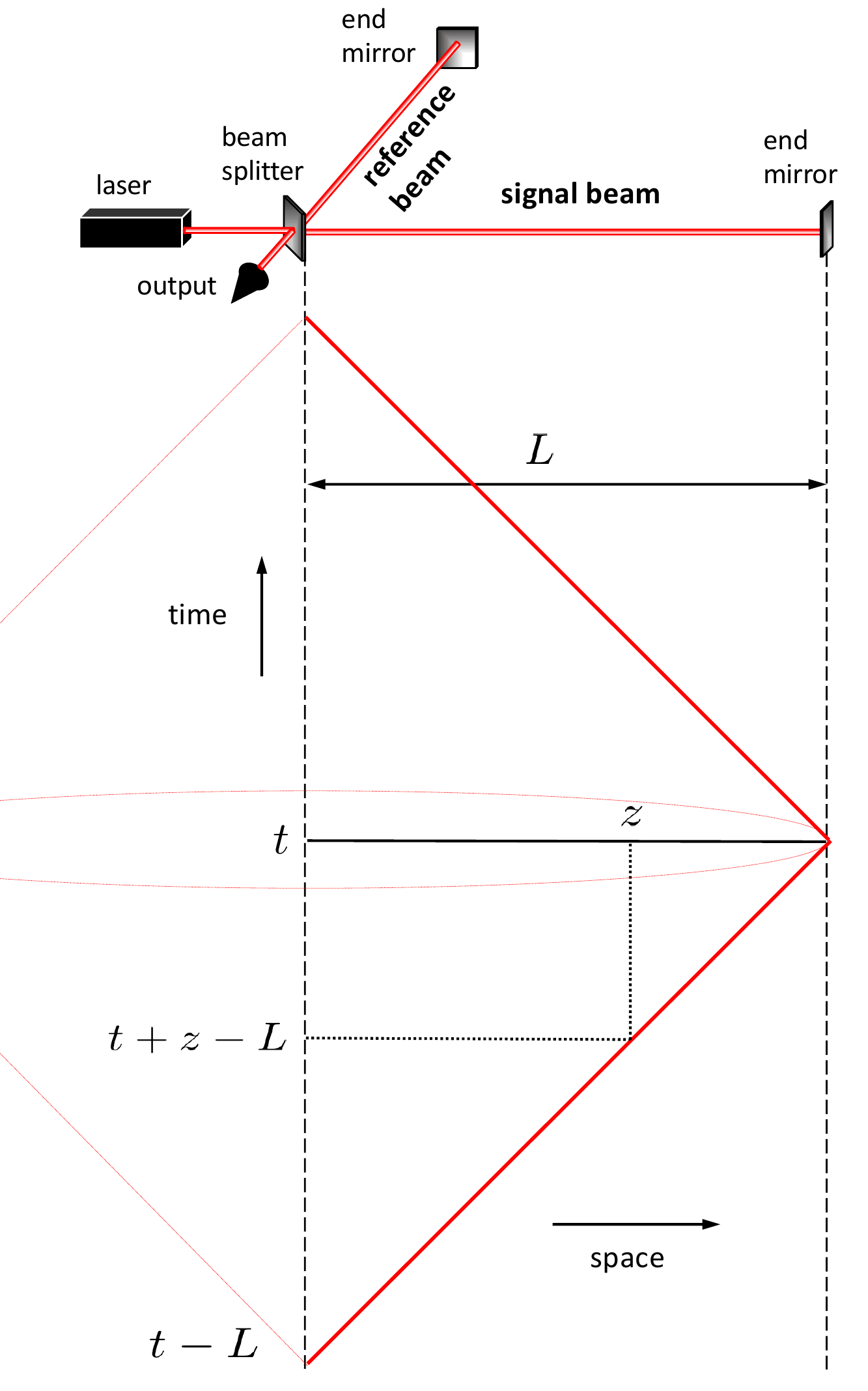}
\caption{Written as a Penrose diagram, the interferometer measures a region of spacetime enclosed by a causal diamond.   On the lower half of the diamond $v = r - t = L$ (with $u$ varying $-L$ to $L$)  while on upper half of the diamond $u = r + t = L$ (with $v$ varying $L$ to $-L$).  The effect can be projected as the quantum uncertainty in the position of the bifurcate horizon where the two rays meet the end mirror at $u = v = L$.  
Figure from Ref.~\cite{VZ1}.}
\label{figure-causal-diamond}
\end{figure}

  A measurement occurs over a finite period of time, during which only part of the spacetime is in causal contact.  Therefore, a measurement defines a light-sheet horizon, shown in Fig.~\ref{figure-causal-diamond}. 
As long as we are interested in only the part of spacetime inside the causal diamond, the metric in some common spacetimes can be mapped to a ``topological black hole''.  If we consider, for example, Minkowski space in four dimensions, we have
\begin{equation}
ds^2 = du dv + dy^2 = - f(R) dT^2 + \frac{dR^2}{f(R)} + r^2 (d\theta^2 + \sin^2 \theta d\phi^2),
\end{equation}
where 
\begin{equation}
f(R) = 1 - \frac{R}{L} + 2 \Phi
\label{eq:blackening}
\end{equation}
and $\langle \Phi \rangle = 0$ in vacuum.  We refer the reader to Ref.~\cite{VZ1} for details of the metric transformation.  This form is known as a topological black hole because it resembles that of a Schwarschild black hole, but with a blackening factor $f(R)$ different from the usual form.  

We now apply the dictionary in Table~\ref{table:dictionary} to causal diamonds in flat empty space, in four dimensions.  Via the dictionary, a spacetime volume has a number of holographic degrees of freedom given by Eq.~(\ref{eq:BH}).  Also via the dictionary, the temperature is set by the size of the horizon
\begin{equation}
T = \frac{f'(R)}{4 \pi} = \frac{1}{4 \pi L}.
\end{equation}
Now these degrees-of-freedom can be treated as bits that fluctuate randomly.  As such they have stochastic energy fluctuations again given by root-N statistics:
\begin{equation}
\Delta M \sim \sqrt{S_{\rm ent}} T = \frac{1}{\sqrt{2} \ell_p}.
\label{eq:MassFlucs}
\end{equation}
Here $S_{\rm ent} = A / 4 G_N$ with $A = 4 \pi L^2$ the area of the spherical entangling surface (shown in the right panel of Fig.~\ref{figure-RW}) and 
\begin{equation}
8 \pi G_N = \ell_p^2
\label{eq:ellp}
\end{equation} 
in four dimensions.  (As discussed below, slightly more formal way of deriving this result is again via the dictionary, $F_\beta = - T S_{\rm ent}$, with $\langle \Delta M^2 \rangle = -\frac{\partial^2}{\partial\beta^2} (\beta F_\beta)$, where $\beta = 1/T$.)  Now this mass sources a Newton potential:
\begin{equation}
\Phi = -\frac{\ell_p^2 \Delta M}{8 \pi L} \sim \frac{\ell_p}{L}. 
\label{eq:NewtPot4d}
\end{equation}
A non-zero potential will shift the location of the horizon, located at $f(R) = 0$ in Eq.~(\ref{eq:blackening}), $\delta L = 2 \Phi L$.  Importantly, a linear shift in the position of the horizon in the topological black hole coordinates is {\em quadratic} in the light-cone coordinates
\begin{equation}
\delta u \delta v = 2 \Phi L^2 \equiv \delta L^2.
\label{eq:lengthfluc4d}
\end{equation}
We thus learn
\begin{equation}
\delta L^2 = \frac{\ell_p L}{4 \pi},
\label{eq:VZ1}
\end{equation}
which is again in accord with the random walk intuition Eq.~(\ref{eq:RW}), as well as the quantum uncertainty of a black hole horizon in $d = 4$ Eq.~(\ref{eq:Marolf4d}), as computed by Marolf with different methods.

In the context of Ryu-Takayanagi causal diamonds in AdS/CFT, a similar argument can be formulated more formally \cite{VZ2}.  The analogue of $\Delta M$ in Eq.~(\ref{eq:MassFlucs}) is fluctuations in the ``modular Hamiltonian,''  
\begin{equation}
K \equiv \int dB^\mu \xi^\nu T_{\mu \nu},
\label{eq:Kdef}
\end{equation}
where $\xi^\nu$ is the conformal Killing vector on the boundary and $dB^\mu$ is the infinitesimal volume element of the boundary CFT at the bifurcate horizon.  Note that Eq.~(\ref{eq:Kdef}) says that modular fluctuations in the vacuum state are vacuum stress tensor fluctuations.  The modular Hamiltonian fixes the density matrix of the boundary CFT:
\begin{equation}
\rho = \frac{e^{-\beta K}}{\mbox{Tr}\left(e^{-\beta K}\right)},
\end{equation}  
with the partition function $Z_\beta$ and free energy $F_\beta$ being defined in the usual way
\begin{equation}
Z_\beta = \mbox{Tr}\left(e^{-\beta K}\right),~~~~~F_\beta = -\frac{1}{\beta} \log Z_\beta.
\end{equation}
The fluctuations of the Modular Hamiltonian $\Delta K$ can then be computed via the free energy:
\begin{eqnarray}
\langle K \rangle & = & \frac{\partial}{\partial\beta} \left(\beta F_\beta\right)\\ \nonumber
\langle \Delta K^2 \rangle & = & - \frac{\partial^2}{\partial\beta^2} \left( \beta F_\beta \right).
\end{eqnarray}
$\langle K \rangle,~\langle \Delta K^2 \rangle$ can be computed holographically via topological black hole thermodynamics \cite{VZ2}, via a bulk geometric calculation~\cite{Nakaguchi:2016zqi,deBoer:2018mzv}, or via the boundary CFT~\cite{perl}.  All calculations give
\begin{equation}
\langle \Delta K^2 \rangle  = \frac{A_\Sigma}{4 G_N},
\label{eq:modflucs}
\end{equation}
where $A_\Sigma$ is the area of the entangling surface.    One can then compute the (fluctuating) gravitational potential sourced by these modular fluctuations.  The result is \cite{VZ2}
\begin{equation}
\langle \Phi^2 \rangle = \frac{\langle \Delta K^2 \rangle}{(d-2)^2} \left(\frac{4 G_N}{A_\Sigma}\right)^2 = \frac{1}{(d-2)^2} \frac{4 G_N}{A_\Sigma},
\label{eq:genNewtPot}
\end{equation}
where $G_N$ is the Newton constant for the bulk $d$-dimensional gravitational theory.  In $d = 4$, identifying $8 \pi G_N = \ell_p^2$, this result agrees with Eq.~(\ref{eq:NewtPot4d}) (up to a factor of $d - 2$ which has to do with how the mass/modular fluctuation was computed in the two cases).  This Newton potential sources length fluctuations similarly to Eq.~(\ref{eq:lengthfluc4d}), which works out to be:
\begin{equation}
\delta L^2  = L^2 \sqrt{\langle \Phi^2 \rangle} =  \frac{L^2}{(d-2)}\frac{1}{\sqrt{S_{\rm ent}}}.
\label{eq:LengthFluctAdS}
\end{equation} 
Setting $S_{\rm ent} \sim L^2 / G_N$ for $d = 4$, we again find a scaling agreeing with the random walk intuition.

\begin{figure}[t]
\includegraphics[scale=0.4]{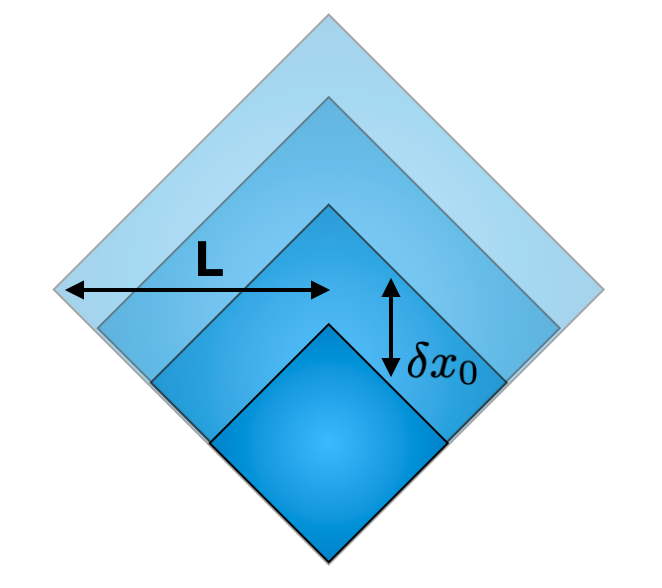}
\caption{Nested Causal Diamonds, each separated by a decoherence scale $\delta x_0 = \tilde \ell_p$.  The decoherence scale is the diffusion constant $D$ in the random walk intuition.  Figure taken from Ref.~\cite{BZ}.}
\label{figure-CD}
\end{figure}

Eqs.~(\ref{eq:genNewtPot})-(\ref{eq:LengthFluctAdS}) give a clue as to how the result Eq.~(\ref{eq:NewtPot4d}) should be generalized to an arbitrary number of dimensions.   Let us return to the nested causal development of a sphere shown in the right panel of Fig.~\ref{figure-RW}.  These nested spheres are depicted as nested causal diamonds in Fig.~\ref{figure-CD}.  According to an argument of Ref.~\cite{BZ}, the subsequent diamonds will become statistically uncorrelated when the change in the entanglement entropy increases as the square-root of the number of degrees of freedom
\begin{equation}
\delta S_{\rm ent} \sim \sqrt{S_{\rm ent}}.
\end{equation}
This implies a scrambling length
\begin{equation}
\delta x_0 \simeq \frac{x_0}{(d-2)} \frac{1}{\sqrt{S_{\rm ent}}}.
\label{eq:scrambling}
\end{equation}
Because these uncertainties in each nested causal diamond are statistically uncorrelated, they will give an accumulated uncertainty
\begin{equation}
\delta L^2 \simeq \delta x_0^2 N = \frac{L^2}{d-2} \frac{1}{\sqrt{S_{\rm ent}}}.
\label{eq:LengthFluctNested}
\end{equation}
Note that this result is in exact agreement with Eq.~(\ref{eq:LengthFluctAdS}), as well as the scaling for the uncertainty in a black hole horizon given by Eq.~(\ref{eq:Marolf}).

Eq.~(\ref{eq:scrambling}) suggests that one should define a fundamental uncertainty scale given by
\begin{equation}
\delta x_0 \equiv \tilde \ell_p = \frac{L}{\sqrt{S_{\rm ent}}}.
\label{eq:tildelp}
\end{equation}
In $d = 4$, $\tilde \ell_ p \sim \ell_p$, but in $d > 4$ this scale is smaller than the Planck length, which we understand to mean that the VZ memory effect does not exist in higher dimensions.  This fundamental uncertainty scale corresponds to the diffusion constant $D$ in the random walk intuition.

\section{Evidence for the Dictionary for Causal Diamonds in Flat Space}

Such an effect as large as Eq.~(\ref{eq:VZ1}) will naturally draw skepticism.  In addition, we know that gauge dependence in gravitational observables can be a tricky issue, as evidenced by its long history in gravitational wave physics.  As such we seek multiple, physically equivalent descriptions.  These descriptions should also demonstrate why we expect the calculations of black hole horizons and light sheet horizons of Ryu-Takayanagi diamonds in AdS to extend to light horizons in flat space. The physically equivalent descriptions that have been explored include:
\begin{itemize}
\item Shifts in Light Ray Operators induced by shockwaves.  One way to think about modular fluctuations $\langle \Delta K^2 \rangle$ is via a series of multiple shockwaves, as shown in Ref.~\cite{VZ3}.  The typical time separation between shocks that a photon experiences while on its trajectory sets the diffusion constant in the random walk picture.  One can show with the 't Hooft uncertainty relations that the relevant length/time scale is given by Eq.~(\ref{eq:tildelp}).
\item An effective description in terms of a bosonic degree of freedom in a high occupation number state \cite{pixellon}.  This model mimics the behavior  expected from a more fundamental effective hydrodynamic description, such as was studied in \cite{Policastro:2002se,Compere:2011dx,deBoer:2015ija,Nickel:2010pr}.  In the context of AdS/CFT, the hydrodynamic mode controls the relative distance between the bifurcate horizon and the boundary~\cite{Crossley:2015tka}, and hence also controls the amount of time required to traverse a causal diamond~\cite{GLZ}.
\item Lower-dimension models reached by a dimensional reduction of the 4-d Einstein Hilbert action in Minkowski space.  This was suggested as a route forward in Ref.~\cite{BZ}, where the 4-d Einstein-Hilbert action in flat empty space was dimensionally reduced to a dilaton theory, following classic work of Carlip and Solodukhin on black hole horizons \cite{carlipreview,solo}.  It was then shown that in the near-horizon limit, the modular fluctuations computed from the dilaton theory, $\langle \Delta K^2 \rangle$, matched those computed for an Ryu-Takayanagi diamond in AdS/CFT, Eq.~(\ref{eq:modflucs}), as well as  for a causal diamond on a flat Randall-Sundrum-II brane.  
\item A description in terms of a saddle point expansion of the gravitational effective action.   Motivated by the conformal description of near-horizon states proposed in Ref.~\cite{BZ}, a partition function from Euclidean effective action of the form
\begin{equation}
Z_\beta = e^{-S_E} = \int dE e^{B \sqrt{E} - \beta E}
\end{equation}  
describes the modular fluctuations 
in Eq.~(\ref{eq:modflucs}).  One can also show that in the near horizon limit, the 4-d Einstein-Hilbert theory can be dimensionally reduced to JT gravity \cite{GLZ}, which shares important features in common with the dilaton theory considered in Ref.~\cite{BZ}.  Since JT gravity can be solved as a QM problem, Ref.~\cite{GLZ} solved for the uncertainty in the light travel time, and reproduced a result consistent with Eq.~(\ref{eq:VZ1}). 
\item Time-Ordered and Out-of-Time-Ordered Correlators of four-point functions of the 't Hooft light ray operators $X^u,~X^v$.  The connection between TOC/OTOCs in the context of JT gravity and multiple shocks have already been explored in the literature, and it remains to make the connection with the observables at hand.  
\end{itemize}
There is a growing web of connections with recent formal advances that suggest that quantum gravity effects could be observable via the connection with the IR scale.

\section{Experimental Tests and Outlook}

What do we seek to test?  It is the {\em fundamental uncertainty} in light ray operators:
\begin{eqnarray}
\delta v(y) = \tilde \ell_p^2  \int _{-L}^L du \int d^{d-2} y' f(y,y') T_{uu}(u,y') \\ \nonumber
\delta u(y) = \tilde \ell_p^2  \int _{-L}^L dv \int d^{d-2} y' f(y,y') T_{vv}(u,y')
\label{eq:lightray} 
\end{eqnarray}
where $\tilde \ell_p$ is given by Eq.~(\ref{eq:tildelp}), $f(y,y')$ is the Green function of the Laplacian on the transverse directions, and $T_{uu},~T_{vv}$ is the stress tensor, due to a vacuum fluctuation, along the $u= r+t,~v = r - t$ direction.  These operators, induced by a non-zero value of the stress tensor generated by a {\em quantum} fluctuation in the vacuum, are evaluated on light sheets at the boundary of the causal diamond.  
Physically, the light ray operators are shifts, $\delta u,~\delta v$ in the position of the light fronts on the upper and lower part of the causal diamond, respectively.   
If we model a vacuum fluctuation as a massless energetic particle carrying momentum $p^v$ at some position $u_i$ along one lightcone direction and transverse position $y_0$, we have
\begin{equation}
T_{uu}(u,y) = p^v \delta(u - u_i) \delta^{(d-2)}(y,y_0).
\end{equation}
We can see from Eq.~(\ref{eq:lightray}), that such a stress tensor on the lower half of the causal diamond would be perpendicular to the light front and cause a shift $\delta v$ in the time of arrival of the light front.  Likewise for the upper half of the the causal diamond, swapping $u \leftrightarrow v$.
These operators satisfy the 't Hooft uncertainty relations \cite{tHooft:1996rdg,tHooft:2018fxg}
\begin{eqnarray}
\langle X^u(y) X^v(y') \rangle = \tilde \ell_p^2 f(y,y'). 
\label{eq:funduncertainty}
\end{eqnarray}
Thus, as a result of Eq.~(\ref{eq:lightray}), we envision fluctuations in the stress tensor (quantified in terms of their two-point, which is directly related to the modular Hamiltonian via Eq.~(\ref{eq:Kdef})) source the fundamental uncertainty in Eq.~(\ref{eq:funduncertainty}).  


As diverse theoretical approaches lead to a consistent picture for a physical effect, the case for experimental tests grows.
The effect in Eq.~(\ref{eq:VZ1}) is small,
numerically corresponding to a signal
\begin{equation}
\sqrt{\langle \delta L^2 \rangle} \approx 8 \times 10^{-18} \mbox{m} \left(\frac{L}{10 \mbox{ m}}\right),
\label{eq:VZnum}
\end{equation}
where we have introduced a factor $\alpha$ to take into account theoretical uncertainties on the model and $\ell_p$ is related to $G_N$ in four dimensions via the conventions in Eq.~(\ref{eq:ellp}); all existing theoretical approach lead to $\alpha = {\cal O}(1)$ as seen explicitly in Eq.~(\ref{eq:LengthFluctAdS}).  
Such a small signal may still be within reach.  This result is integrated over a the spacetime volume enclosed by the interferometer, as shown in the right panel of Fig.~\ref{figure-RW}.  To have a fully inclusive phenomenological prediction, one must have both the angular correlations (to determine the relative strength of the signal between the interferometer arms) and the time dependence of the signal ({\em i.e.} the power spectral density).  

This was a motivation for the ``pixellon'' model presented in Ref.~\cite{pixellon}.  The idea here was to treat the mass/modular fluctuations represented in Eqs.~(\ref{eq:MassFlucs}),~(\ref{eq:modflucs}) as a scalar fluid (the pixellon) with a high density of states given by $\Delta K / K$, where $\Delta K = \sqrt{\langle \Delta K^2 \rangle}$.  The pixellon was then coupled to mirrors gravitationally, and the Feynman-Vernon influence functional utilized to compute fluctuations in the position of a mirror.  This model allowed to compute the power-spectral-density, but not yet the angular correlations in two interferometer arms separated by some angle.

Since the vacuum fluctuations are uncorrelated on timescales longer than a single light-crossing time, an instrument like LIGO has weakened sensitivity to these effects since a typical photon remains in the Fabry-Perot cavity for dozens of light crossing-times, over which period the signal is averaged down.  Instead, a simple Michelson interferometer, where the light sheets traverse a single causal diamond will give rise to a potentially observable signal.  There are several instruments that are of this class, but none yet with strong enough sensitivity to definitively probe the VZ effect.  For example, as discussed in Ref.~\cite{pixellon}, the Fermilab Holometer's \cite{Chou:2017zpk} sensitivity is a factor of a few above that needed.

An experiment targeting the VZ effect has been proposed, Gravity from the Quantum Entanglement of SpaceTime (GQuEST) \cite{GQuest}, a 5 m instrument aiming to reach a strain sensitivity.  The fluctuation in Eq.~(\ref{eq:VZ1}) corresponds approximately to a strain autocorrelation
\begin{equation}
C_h(\tau,\theta) \equiv \frac{\ell_p}{(4 \pi)^2 L} g(\tau,\theta)
\end{equation}
where $g(\tau,\theta)$ is a form factor that depends on the angle $\theta$ between the two interferometer arms and the time separation $\tau$ between measurements of the arm length difference; $g(\tau,\theta)$ is expected to be ${\cal O}(1)$ for time separations on the order of L and angular separations ${\cal O}(1)$~\cite{CLZ}.  The power spectral density in peaked at $f_{\rm peak} \approx c / 4 L$ with a width $\Delta F_{\rm sig} \approx c / 2 L$.  
GQuEST is designed to reach $g(0,\pi/2) = 1$ at $3-\sigma$ sensitivity, at peak signal frequency of 15 MHz, with a 5 m arm and 15 kW of laser power on the beamsplitter, after 1000 s of integration time.  Other experiments, such as Ref.~\cite{Vermeulen:2020djm}, are aiming for strain sensitivity on the same order of magnitude with a simple Michelson interferometer.

The coming years promise both a theoretically and experimentally active period of development, in search for spacetime fluctuations from quantum gravity.  A more active interplay between theory and experiment on the question of quantum gravity will stimulate the use of known tools for new questions.  Experiments sharpen the theoretical mind and point the way towards new opportunities.  We seek to bridge the divide between quantum mechanics and gravity, in the laboratory, by exploring the quantum nature of horizons.  

\subsection*{Acknowledgments}

I thank my collaborators Erik Verlinde, Tom Banks, Yanbei Chen, Dongjun Li, Vincent Lee, Sergei Gukov, Lee McCuller, Rana Adhikari, Cynthia Keeler, Temple He and Allic Sivaramakrishnan for ongoing discussion and work on these directions. This is supported by the Heising-Simons Foundation ``Observational Signatures of Quantum Gravity'' collaboration grant 2021-2817, by the DoE under contract DE-SC0011632, and by a Simons Investigator award.

\newpage 

\bibliography{QG}

\end{document}